\documentclass{emulateapj}

\def \kms{$\rm{km\ s^{-1}}$}

\def \pow10#1{\times 10^{#1}}

\journalinfo{The Astronomical Journal, 000:L000-L000, 2009}
\slugcomment{Accepted for publication in \aj}

\tighten
\begin{document}

\title{The Frequency of Tidal Features Associated with Nearby Luminous
  Elliptical Galaxies from a Statistically Complete Sample}

\author{Tomer Tal\altaffilmark{1},
  Pieter G. van Dokkum\altaffilmark{1},
  Jenica Nelan\altaffilmark{1},
  Rachel Bezanson\altaffilmark{1}}
\altaffiltext{1}{Yale University Astronomy Department, P.O. Box
  208101, New Haven, CT 06520-8101 USA}

\shorttitle{The Assembly Rate of Nearby Ellipticals}
\shortauthors{Tal et al.}

\begin{abstract}
  We present a deep broadband optical imaging study of a complete
  sample of luminous elliptical galaxies ($M_B<-20$) at distances 15
  Mpc - 50 Mpc, selected from the Tully catalog of nearby galaxies.
  The images are flat to $\sim$0.35\% across the 20' field and reach
  a V band depth of 27.7 mag arcsec$^{-2}$.
  We derive an objective tidal interaction parameter for all galaxies
  and find that 73\% of them show tidal disturbance signatures in
  their stellar bodies.
  This is the first time that such an analysis is done on a
  statistically complete sample and it confirms that tidal features in
  ellipticals are common even in the local Universe.
  From the dynamical time of the sample galaxies at the innermost
  radius where tidal features are detected we estimate the mass
  assembly rate of nearby ellipticals to be $dM/M\sim0.2$ per Gyr with
  large uncertainty.
  We explore the relation between gravitational interaction signatures
  and the galaxy environment and find that galaxies in clusters are
  less disturbed than group and field galaxies.
  We also study how these interactions affect the broadband colors of 
  ellipticals and find a moderate correlation, suggesting that the
  mergers are not accompanied by significant star-formation.
  Lastly, we find no correlation between AGN activity, as
  measured by 6cm radio emission, and large scale tidal distortions.
  This implies that gravitational interactions are not the only, and
  perhaps not the most important, trigger of nuclear activity.
  In summary, we find that elliptical galaxies in groups and low
  density environments continue to grow at the present day through
  mostly ``dry'' mergers involving little star formation.
\end{abstract}

\keywords{
galaxies: interactions  ---
galaxies: evolution ---
galaxies: elliptical ---
galaxies: structure}
\section {\label{intro}Introduction}
  Giant elliptical galaxies dominate the high end of the mass spectrum
  in the nearby Universe and are mostly composed of old stellar
  populations, in contradiction to na\"{i}ve predictions from
  hierarchical growth models.
  In a $\Lambda$CDM Universe the most massive dark matter halos form
  later than less massive ones and the elliptical galaxies that
  reside in them are therefore expected to have formed recently
  \citep[e.g.,][]{kauffmann_formation_1993}.
  This, however, does not agree with photometric studies of
  ellipticals that produce red colors in all of these galaxies,
  implying their stars are mostly old 
  \citep[e.g.][]{visvanathan_color-absolute_1977, bower_precision_1992,
  chang_colours_2006}.
  The lack of recent star formation is further supported by results
  from HI surveys of early-type galaxies that find extremely low cold
  gas content in these systems when compared to spiral galaxies
  \cite[e.g.][]{sanders_neutral_1980, knapp_statistical_1985,
  sadler_hi_2001}.

  A possible solution may be that galaxies grow through ``dry''
  mergers which do not lead to star
  formation \citep{kormendy_recognizing_1984, white_galaxy_1991,
  kauffmann_age_1996, dokkum_recent_2005, naab_properties_2006,
  boylan-kolchin_red_2006}.
  The little gas that is accreted into the elliptical by the
  interaction may be heated and kept from cooling by nuclear feedback
  \citep{croton_many_2006} or the gravitational interactions
  themselves \citep{dekel_gravitational_2008,
  kenney_spectacular_2008}.
  
  One of the most direct ways to study the assembly rate of elliptical
  galaxies is to measure the incidence of tidal features.
  \cite{schweizer_correlations_1992} studied a sample of 69 early-type
  galaxies with a recession velocity of less than 4000 kms$^{-1}$.
  The galaxies were selected from a parent sample of 145 objects
  to exclude most cluster members and to include a few galaxies that
  had been previously known as interacting systems.
  Optical colors were measured from the sample and a ``fine
  structure'' parameter was assigned to each galaxy by visually
  characterizing the amount and type of their morphological
  disturbances.
  In their paper, \citeauthor{schweizer_correlations_1992} found
  that the broadband colors of early-type galaxies are correlated
  with their respective fine structure parameter and therefore also
  with gravitational interactions.

  A different approach to quantifying the morphological disturbances
  of the stellar bodies of ellipticals was performed by
  \cite{dokkum_recent_2005} [hereafter vD05], who studied a sample of
  126 early-type galaxies at a median redshift of 0.11.
  vD05 divided deep broad-band optical images of the galaxies by their
  respective fitted smooth stellar models and measured the residuals
  of the quotients.
  This quantitative approach facilitates comparison to simulations
  \citep[e.g.][]{feldmann_tidal_2008} and other data
  \citep[e.g.][]{mcintosh_ongoing_2008}.

  Here we perform, for the first time, a quantitative analysis of
  morphological disturbances in a complete sample of elliptical
  galaxies at z$\approx$0.
  The field of view of current generation imagers, along with
  excellent flat fielding, allow us to study the stellar bodies of
  nearby ellipticals and to constrain the rate of gravitational
  interaction in the nearby Universe.
  The results of this study can be used in numerical simulations and
  theoretical models of galactic evolution to constrain the
  formation and mass growth of these objects at recent cosmic times.
  Furthermore, most of our sample is composed of well-known
  bright galaxies that have been studied extensively in the past.
  
    \begin{figure}
      \includegraphics[width=0.47\textwidth]{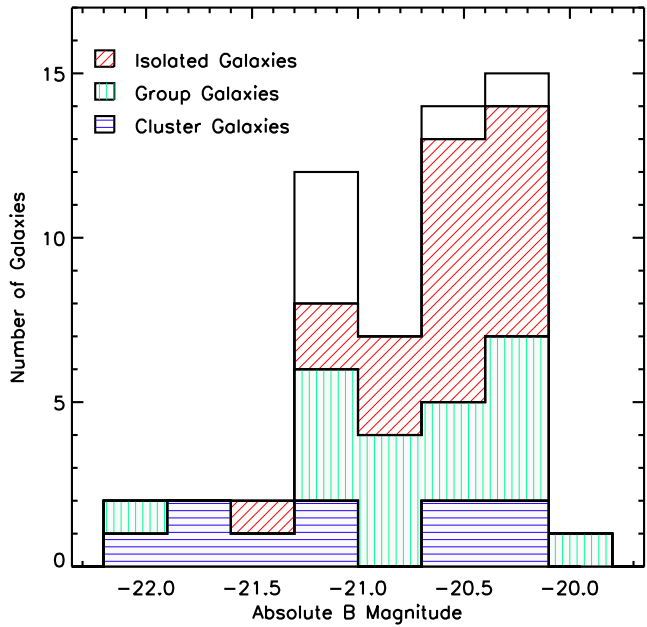}
      \caption{The luminosity and environment distributions of the 55
        sample galaxies.}
      \hfill
      \label{fig:sample_dist}
    \end{figure}

\section{\label{data}Data}
  Observations of a volume limited sample of luminous elliptical
  galaxies were obtained with the Y4KCam imager mounted on the SMARTS
  1m telescope at CTIO in four epochs between 2006 and 2008.
  The instrument is a 4Kx4K CCD camera optimized for wide-field
  broad-band imaging, providing a nearly undistorted 20'x20' field of
  view.
  
  \subsection{Sample Selection}
    Candidates for inclusion in this study were initially selected
    from the Nearby Galaxies Catalog \citep{tully_nearby_1988},
    consisting of all elliptical galaxies at a declination between -85
    and +10.
    A distance cut was applied to the initial sample to exclude
    galaxies that are farther than 50 Mpc or closer than 15 Mpc.
    The distance threshold ensures that the outer parts of all
    galaxies fit in the instrument's field-of-view and that a
    sufficient signal-to-noise ratio is achieved for all targets.
    A luminosity cut of $M_B<$-20.15 was used, with the magnitudes
    taken from \cite{tully_nearby_1988} and converted to our
    cosmology.
    This limit corresponds to $M_B<$-20.0 in the
    \cite{tully_nearby_1988} atlas, as he used $H_0=75$ km s$^{-1}$
    Mpc$^{-1}$, and to L$\geq$L$_{\ast}$ \citep{blanton_galaxy_2003}.
    The last selection criterion that we used rejected galaxies with
    Galactic latitude of less than 17 due to difficulties of
    constraining a good model fit in a crowded stellar field.  

    One galaxy (NGC 5796) was excluded due to poor observing
    conditions that resulted in noise levels over 5 times worse than
    the sample mean.
    The final catalog therefore consists of 55 giant ellipticals
    (figure \ref{fig:sample_dist}) and it includes members of four
    nearby clusters (Virgo, Fornax, Centaurus and Antlia).
    Galaxy environments were determined from the literature using
    NASA's Astrophysics Data System Bibliographic Services to find
    references for previous studies of the objects.
    The methods used to derive these assumed environments are
    therefore inconsistent throughout the sample and they vary in
    accuracy.
    
    \begin{figure}
      \includegraphics[width=0.47\textwidth]{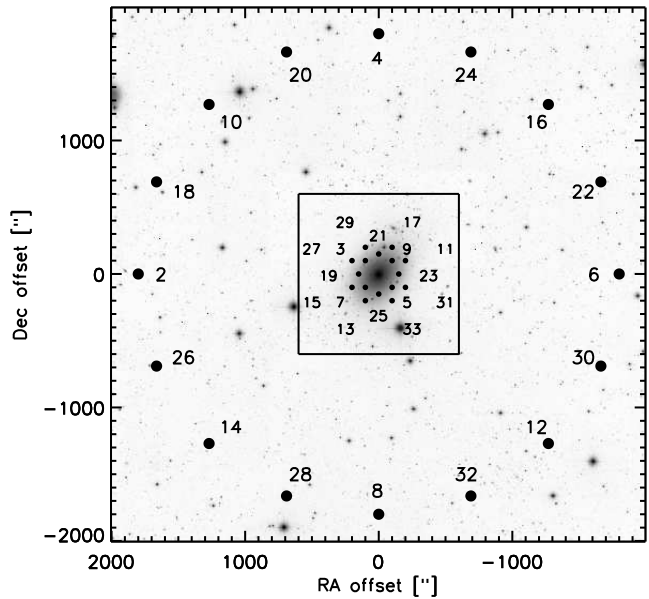}
      \caption{The observing pattern of a galaxy at (0,0) overlaid on
      an SDSS image of NGC 4636.
      Filled circles represent telescope pointings for both target and
      sky frames, which are accompanied by a number denoting their
      place in the sequence.
      The central square shows the angular size of the field of view.}
      \hfill
      \label{fig:obspat}
    \end{figure}
  
    \begin{figure*}
      \includegraphics[width=0.98\textwidth]{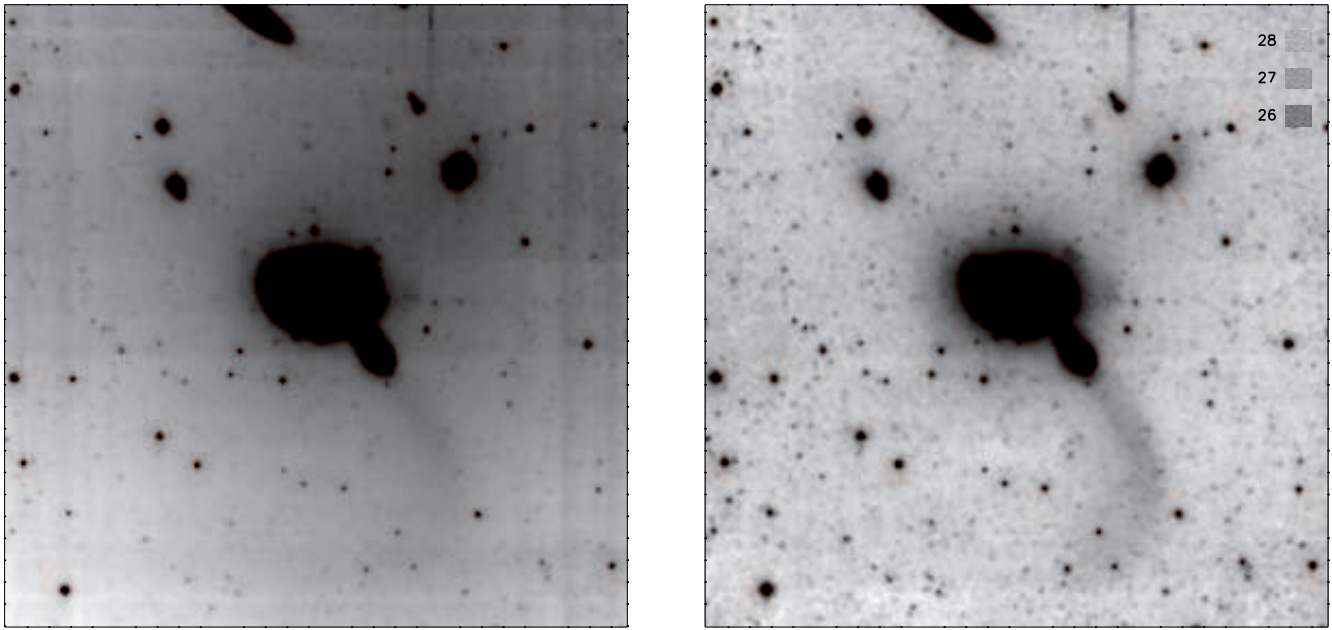}
      \caption{A demonstration of the importance of a high degree of
      field flatness for studying faint interaction signatures.
      The same data were reduced with and without the use of a
      dark-sky flat field frame (left and right panels,
      respectively).
      Note that in the left panel the tidal tail extending away from
      NGC 5576 cannot be distinguished from flattening
      distortions in the field. The three boxes in the upper-right
      corner have surface brightness values similar to the faintest
      detectable features. Both images were further smoothed and
      binned for presentation purposes.}
      \hfill
      \label{fig:flatness}
    \end{figure*}
   
\begin{table*}[ht]
  \caption{Catalog}
  \centering
  \begin{tabular}{l c c c c c c l}
    \hline\hline
    Object & R.A\footnotemark[a] & Dec.\footnotemark[a] & M$_B$
    \footnotemark[b] & $(B-V)$\footnotemark[c] & Env.\footnotemark[d]
    & T$_{c}$ \footnotemark[e] & Tidal Feature Description\\
    \hline
    NGC 584 & 01:31:20.7 & -06:52:05 &  -20.83 & 0.92 & F & 0.076 &
    Non-spherical isophotes; Spiral companion\\
    NGC 596 & 01:32:52.1 & -07:01:55 & -20.37  & 0.90 & F & 0.110 &
    Pronounced outer shell\\
    NGC 720 & 01:53:00.5 & -13:44:19 & -20.53  & 0.96 & F & 0.079 &
    Boxy outer isophotes\\
    NGC 1199 & 03:03:38.4 & -15:36:49 & -20.24 & 0.97 & G & 0.067 &
    Intra-group diffuse emission\\
    NGC 1209 & 03:06:03.0 & -15:36:41 & -20.51 & 0.95 & G & 0.116 &
    X-like isophotal structure; NW linear feature\\
    NGC 1399 & 03:38:29.1 & -35:27:03 & -20.49 & 0.98 & C & 0.064 &
     \\
    NGC 1395 & 03:38:29.8 & -23:01:40 & -20.44 & 0.94 & F & 0.094 &
    NW perpendicular feature\\
    NGC 1407 & 03:40:11.9 & -18:34:49 & -21.17 & 0.93 & G & 0.083 &
     \\
    NGC 2865 & 09:23:30.2 & -23:09:41 & -20.85 & 0.78 & F & 0.193 &
    Multiple shells and tail-like features \\
    NGC 2974 & 09:42:33.3 & -03:41:57 & -20.78 & 0.95 & F & 0.110 &
    Multiple shells\\
    NGC 2986 & 09:44:16.0 & -21:16:41 & -20.85 & 0.99 & F & 0.045 &
    Possibly paired with disturbed spiral \\
    NGC 3078 & 09:58:24.6 & -26:55:37 & -20.78 & 0.97 & F & 0.103 &
     \\
    NGC 3258 & 10:28:53.6 & -35:36:20 & -20.47 & 0.92 & G & 0.123 &
    Possibly paired with low mass elliptical\\
    NGC 3268 & 10:30:00.6 & -35:19:32 & -20.43 & 0.96 & C & 0.087 &
    Possibly paired with disturbed spiral\\
    NGC 3557B & 11:09:32.1 & -37:20:59 & -20.28 & 0.86 &  N & 0.182 &
    Isophotal twisting; Asymmetric outer isophotes\\
    NGC 3557 & 11:09:57.6 & -37:32:21 & -22.33 & 0.87 & G & 0.111 &
    SW fan; Asymmetric outer isophotes\\
    NGC 3585 & 11:13:17.1 & -26:45:18 & -21.08 & 0.91 & F & 0.048 &
    Asymmetric outer isophotes\\
    NGC 3640 & 11:21:06.8 & +03:14:05 & -20.93 & 0.92 & F & 0.142 &
    Highly disturbed stellar body\\
    NGC 3706 & 11:29:44.4 & -36:23:29 & -21.29 & 0.93 & U & 0.120 &
    Inner shell\\
    NGC 3904 & 11:49:13.2 & -29:16:36 & -20.26 & 0.94 & U & 0.108 &
    Outer shell\\
    NGC 3923 & 11:51:01.8 & -28:48:22 & -21.38 & 0.95 & U & 0.100 &
    Multiple outer shells\\
    NGC 3962 & 11:54:40.1 & -13:58:30 & -20.88 & 0.95 & F & 0.059 &
     \\
    NGC 4105 & 12:06:40.8 & -29:45:37 & -20.51 & 0.87 & F & 0.109 &
    Interacting with a tidally disturbed spiral\\
    NGC 4261 & 12:19:23.2 & +05:49:31 & -21.56 & 0.98 & C & 0.053 &
    NW tidal arm; Faint SE fan\\
    NGC 4365 & 12:24:28.2 & +07:19:03 & -20.67 & 0.97 & C & 0.070 &
    Faint SW fan\\
    IC 3370 & 12:27:37.3 & -39:20:16 & -21.38  & 0.89 &  U & 0.192 &
    X-like isophotal structure; Broad N fan\\
    NGC 4472 & 12:29:46.7 & +08:00:02 & -21.97 & 0.97 & C & 0.000 &
     \\
    NGC 4636 & 12:42:49.9 & +02:41:16 & -20.83 & 0.93 & F & 0.066 &
     \\
    NGC 4645 & 12:44:10.0 & -41:45:00 & -21.03 & 0.95 & C & 0.000 &
     \\
    NGC 4697 & 12:48:35.9 & -05:48:03 & -21.82 & 0.92 & C & 0.091 &
    Non-spherical inner isophotes\\
    NGC 4696 & 12:48:49.3 & -41:18:40 & -22.20 & 0.94 & C & 0.075 &
    Faint outer shell\\
    NGC 4767 & 12:53:52.9 & -39:42:52 & -21.28 & 0.93 & C & 0.000 &
    Faint inner shell\\
    NGC 5011 & 13:12:51.8 & -43:05:46 & -21.25 & 0.89 & U & 0.077 &
     \\
    NGC 5018 & 13:13:01.0 & -19:31:05 & -21.61 & 0.85 & F & 0.184 &
    Highly disturbed; Multiple tidal tails and shells\\
    NGC 5044 & 13:15:24.0 & -16:23:08 & -21.16 & 0.98 & G & 0.041 &
     \\
    NGC 5061 & 13:18:05.1 & -26:50:14 & -21.34 & 0.85 & F & 0.104 &
    Pronounced tidal tail; Shell\\
    NGC 5077 & 13:19:31.7 & -12:39:25 & -20.67 & 0.98 & G & 0.061 &
     \\
    NGC 5576 & 14:21:03.7 & +03:16:16 & -20.55 & 0.88 & F & 0.122 &
    Interacting pair; tidal tail longer than 75 kpc\\
    NGC 5638 & 14:29:40.4 & +03:14:00 & -21.27 & 0.94 & F & 0.036 &
     \\
    NGC 5812 & 15:00:55.7 & -07:27:26 & -20.73 & 0.94 & F & 0.080 &
    Interacting with dwarf companion; Tidal tail\\
    NGC 5813 & 15:01:11.2 & +01:42:07 & -20.92 & 0.95 & G & 0.054 &
     \\
    NGC 5846 & 15:06:29.3 & +01:36:20 & -21.31 & 0.98 & G & 0.068 &
    Faint outer shells\\
    NGC 5898 & 15:18:13.5 & -24:05:53 & -20.64 & 0.92 & G & 0.114 &
    Three spiral arm-like tidal tails\\
    NGC 5903 & 15:18:36.5 & -24:04:07 & -21.03 & 0.89 & G & 0.075 &
    Possible dust extinction S of the center\\
    IC 4797 & 18:56:29.7 & -54:18:21 & -20.90  & 0.92 & G & 0.226 &
    Tidal tails; SE diffuse emission\\
    IC 4889 & 19:45:15.1 & -54:20:39 & -20.70  & 0.88 & U & 0.158 &
    Isophotal twisting; E fan\\
    NGC 6861 & 20:07:19.5 & -48:22:13 & -20.95 & 0.95 & G & 0.123 &
    Non-spherical isophotes; Intra-group emission\\
    NGC 6868 & 20:09:54.1 & -48:22:46 & -21.21 & 0.97 & G & 0.096 &
     \\
    NGC 6958 & 20:48:42.6 & -37:59:51 & -20.68 & 0.86 & C & 0.122 &
    Multiple shells\\
    NGC 7029 & 21:11:52.0 & -49:17:01 & -20.26 & 0.86 & G & 0.085 &
    Boxy inner isophotes\\
    NGC 7144 & 21:52:42.4 & -48:15:14 & -20.51 & 0.91 & G & 0.100 &
     \\
    NGC 7196 & 22:05:54.8 & -50:07:10 & -20.47 & 0.91 & G & 0.171 &
    Shell\\
    NGC 7192 & 22:06:50.1 & -64:18:58 & -20.70 & 0.92 & F & 0.096 &
    Shell\\
    IC 1459 & 22:57:10.6 & -36:27:44 & -20.73  & 0.96 & G & 0.137 &
    Multiple shells\\
    NGC 7507 & 23:12:07.6 & -28:32:23 & -20.36 & 0.94 & F & 0.084 &
    Faint N shell\\
    \hline\hline
  \end{tabular}
  \footnotetext[a]{J2000.0}
  \footnotetext[b]{Taken from \cite{tully_nearby_1988} and corrected
  to H$_0$=70 km s$^{-1}$ Mpc$^{-1}$}
  \footnotetext[c]{Colors within effective radius from
  \cite{michard_near_2005} and \cite{de_vaucouleurs_third_1991}}
  \footnotetext[d]{Environment: C=cluster; G=group; F=field; U=unknown}
  \label{table:catalog}
  \footnotetext[e]{Corrected tidal parameter as derived in subsection
  \ref{tpderiv}}
\end{table*}				

  \subsection{Observations\label{observations}}
    Broadband optical observations of the sample galaxies had
    been acquired by multiple authors, showing
    some of the tidal features that we discuss in this work.
    However, the standard techniques used to obtain and reduce these
    previous data were not aimed at revealing the faint gravitational
    interaction signatures and they were washed out by residual
    background level variations across the field.
    In order to overcome this we supplemented each
    set of galaxy observations with a sequence of dark sky exposures
    of equivalent depth.
    
    The galaxies were observed in a sequence of 33
    pointings, including both object and dark sky frames.
    We used the V band as it provides the highest signal-to-noise
    ratio in a given exposure time.
    The observing pattern, shown in figure \ref{fig:obspat},
    included seventeen 300 sec exposures of the target galaxy and
    sixteen 300 sec exposures of background sky, acquired in
    alternating order.
    All object frames were shifted from each other by 1-3' in order
    to correct for cosmetic defects in the CCD.
    Figure \ref{fig:obspat} also shows the order in which sky frames
    were observed, collecting data from different sides of the
    galaxy in a semi-random pattern.
    All galaxies were observed with total exposure times between
    4200 and 7200 seconds. 

    In order to improve the signal-to-noise ratio the data were binned
    by a factor 2$\times$2 at the telescope, producing a pixel size of
    0.578''.
    To further increase the sensitivity to tidal features we also
    binned in software, resulting in a final pixel size of 1.156''.
    The typical stellar FWHM of the images used for analysis is
    $\sim$1.7''.

   \subsection{Reduction}
    Initial reduction of the data followed standard techniques and
    consisted of zero level subtraction and first order field
    variation corrections using dome flat frames.

    In order to flatten the field to a higher degree dark-sky
    flat-field frames were prepared for each of the sample
    galaxies.
    All sources were masked in the individual dark-sky exposures which
    were then averaged and applied to the object frames.
    Lastly, the reduced object exposures were aligned and combined to
    create the final data products.

    Apparent magnitudes were calibrated using aperture photometry of
    \cite{prugniel_general_1998} and were corrected for Galactic
    reddening using infrared dust maps from \cite{schlegel_maps_1998}.
    We assume distance measurements from the Tully catalog
    (corrected to our cosmology) to convert the luminosity profiles to
    physical units.

    The data were reduced using the NOAO/IRAF software and are
    available for download at http://www.astro.yale.edu/obey

  \subsection{\label{depth}Depth and Flatness of the Field}
    Our sky flat-fielding procedure (discussed in subsection
    \ref{observations}) resulted in a dramatic improvement to the
    flatness of the field, as can be seen in figure
    \ref{fig:flatness}.
    In this figure the $\sim$85 kpc long tidal tail that extends away
    from the companion of NGC 5576 is barely detectable in the image
    on the left which uses standard flat-fielding but is clearly visible
    in the dark-sky flat-fielded frame.
    
    We used two methods to determine the depths of the sky flat field
    images.
    First, we reduced and stacked a dark sky image in the same way
    that was used to produce the target frames and smooth it to the
    scale of a typical tidal feature ($\sim$20'').
    Although the resultant image was inherently flat, it preserves the
    photon noise level as limited by the telescope, instrument and
    the observing program.
    Second, we aggressively masked a target frame for bright objects
    and subsequently smooth it using a median kernel.
    This frame is insensitive to any pixel-to-pixel variation but it
    reflects the large-scale variation due to residual flattening
    issues or sky conditions at the time of observation.
    We then measured the standard deviation across both frames to
    obtain the detection threshold of faint tidal features.
    For a 20''$\times$20'' box (corresponding to a typical tidal
    feature scale) we derived a 1$\sigma$ photon-noise detection
    threshold of $\sim$29 mag.
    From the flatness limited frame we measured a 1$\sigma$
    detection threshold of $\sim$27.7 mag.
    This result implies that a similar program carried on a larger
    telescope will not necessarily yield a lower detection threshold
    as the data are dominated by residual flatness variations rather
    than photon noise.

\section{Tidal Features}
  \subsection{Morphological Disturbances\label{mordist}}
    A visual inspection of the 55 sample galaxies reveals an
    extraordinary assortment of stellar morphology disturbances in a
    large number of galaxies.
    Similar disturbances have been reported by several authors and
    they can be divided into four main categories:

    \begin{figure*}
      \includegraphics[width=0.98\textwidth]{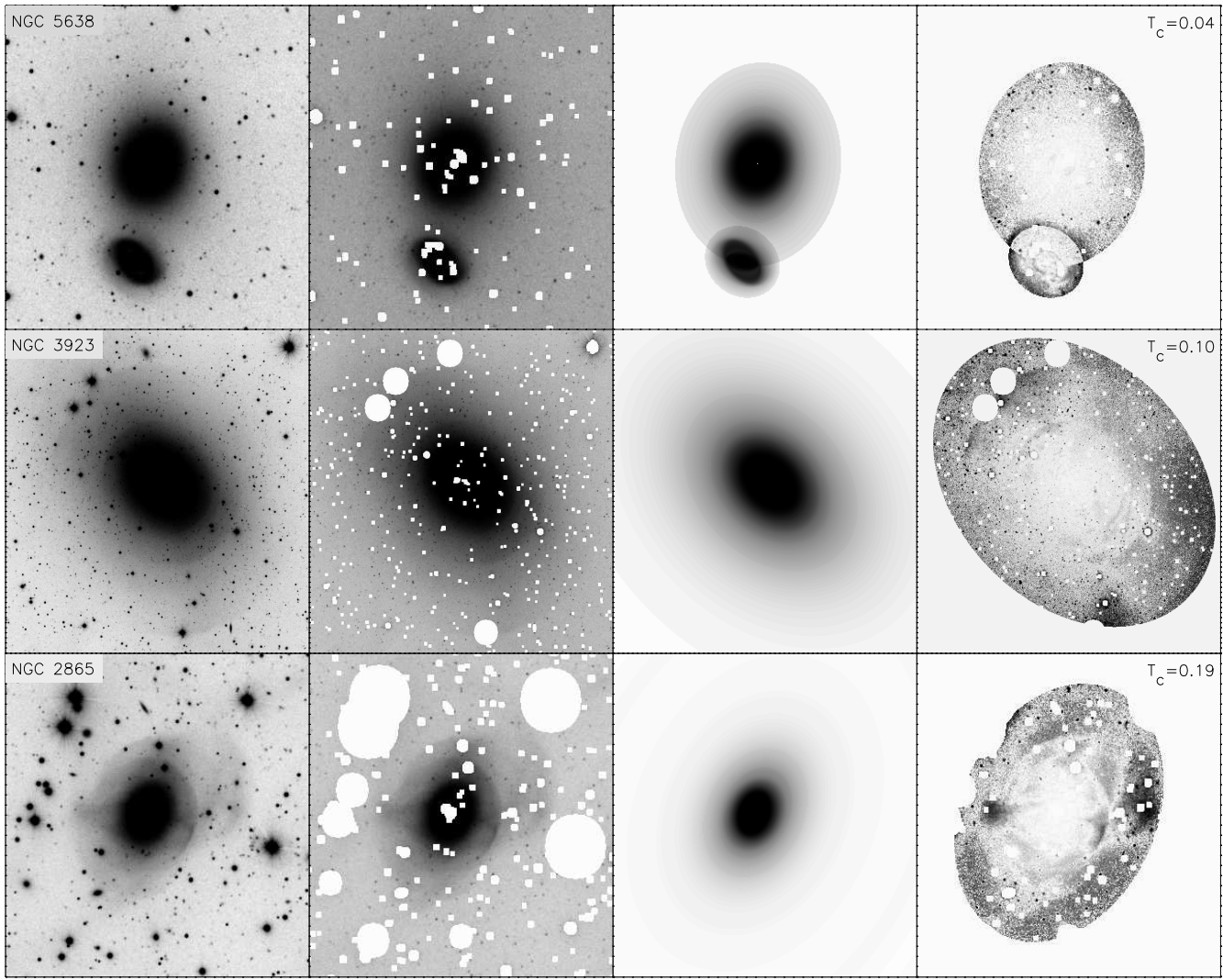}
      \caption{Three examples of the model fitting and division method
      we employ to derive the tidal parameter of the sample
      galaxies.
      From top to bottom: the galaxy NGC 5638 and its companion show
      no signs of gravitational interaction;
      NGC 3923 has shells at multiple radii;
      NGC 2865 is a recent or ongoing merger showing multiple
      interaction signatures.
      From left to right: The total counts image;
      foreground and background object masked frame;
      elliptical model fit to the data;
      the division of the data by the model - this image was used to
      measure the tidal parameter $T$.}
      \hfill
      \label{fig:atlas_sub}
    \end{figure*}
    
    \textbf{Shells} - observed in the stellar component of many nearby
    ellipticals, shells are usually considered
    to result from an accretion of a small companion by a massive
    galaxy.
    Numerical simulations of such interactions have successfully 
    reproduced the observed properties of shells even without the
    inclusion of gas dynamics and star formation
    \citep[e.g.][]{dupraz_shells_1986}.
    In our sample we identify at least 12 galaxies with shells, making
    them the most common interaction signature.

    \textbf{Tidal Tails} - these linear streams of stellar matter are
    less commonly observed than shells in elliptical systems and are
    generally accepted as evidence for a dynamically cold component in
    the accreted companion.
    Tidal tails can also be reproduced in simulations of dry mergers
    when one of the progenitors is rotating \citep{combes_stellar_1995}.
    Tails that are produced in this manner appear
    wide and short-lived in comparison to the long and narrow tails
    produced by interacting spirals.
    An interaction between an elliptical galaxy and a gas-rich disk
    system can also produce such features \citep{feldmann_tidal_2008}.    
    In this work we identify both kinds of tails in a range of widths
    and lengths, some extending as far as 11' ($\sim$85kpc) in
    projected length.

    \textbf{Broad Fans of Stellar Light} - also well reproduced in
    simulations of dry mergers, extended stellar fans typically have a
    low surface brightness which makes them hard to detect in shallow
    surveys.    
    Studies of interacting ellipticals
    \citep[vD05,][]{mcintosh_ongoing_2008} show multiple
    examples of this morphological signature and associate them with
    interactions of gas poor systems.
    This interaction signature is visible in a number of systems in
    our sample, most of which do not have an obvious companion and are
    therefore likely the result of a major merger in the past (see
    figure 1 of vD05).

    \textbf{Highly Disturbed Galaxies} - at least three galaxies
    ($\sim$5\%) in our sample of 55 exhibit signs of an ongoing merger
    that is violently disturbing their stellar components (NGC 2865, NGC
    3640 and NGC 5018).
    These systems show signs of one or more of the above interaction
    signatures, including well defined linear filaments that extend
    radially from the galactic centers.
    Two of the three galaxies are recent merger remnants and in
    the third the interaction is still ongoing with a nearby companion.
    These complex systems probably result from mergers with
    multi-component objects such as S0 or spiral galaxies.\\

    Morphological disturbances in the stellar bodies of the sample
    galaxies are common and are clearly apparent in at least
    two-thirds of the objects.
    The existence and frequency of these features imply that
    elliptical galaxies are still evolving in the nearby Universe
    through collisions and gravitational interactions that affect
    their stellar bodies.
    
    A brief description of the visible distortions of the sample
    galaxies is given in the last column of table
    \ref{table:catalog}.
         
    \begin{figure}
      \includegraphics[width=0.47\textwidth]{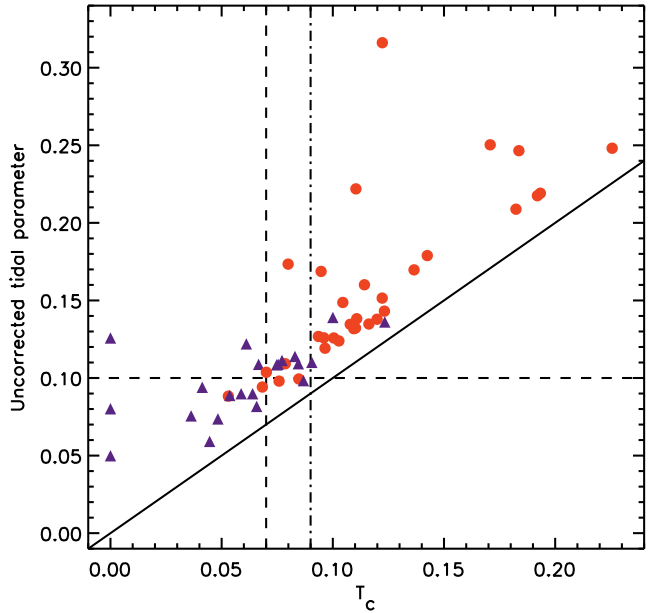}
      \caption{A comparison between the initial and the corrected
      tidal parameter values of the sample galaxies. The vertical
      dashed and dot-dashed lines show the tidal parameter threshold
      for marginally and clearly interacting systems,
      respectively. The horizontal dashed line is shown for comparison
      with vD05 and marks their interaction threshold.
      A visual inspection of the data separates galaxies with detected
      tidal features (red circles) from visually appearing relaxed
      systems (blue triangles).}
      \hfill
      \label{fig:tp_clust}
    \end{figure}
   
  \subsection{Tidal Parameter Derivations \label{tpderiv}}
    Quantitative analysis of the tidal disturbance of the galaxies was
    carried out following the method described by
    \cite{colbert_optical_2001} and vD05.
    Each galaxy was fitted with an elliptical galaxy model using the
    \textit{ellipse} task in IRAF in three iterations, allowing the
    center position, position angle and ellipticity to vary with
    radius.
    Foreground stars and background galaxies were masked using
    \textit{SExtractor}, a source detection software
    \citep{bertin_sextractor:_1996}.
    Masked object frames were then divided by the galaxy models
    and were further binned by a 3$\times$3 factor in order to
    increase the signal-to-noise ratio.
    The resulting tidal parameter is given by:
    \begin{equation}
      T_{galaxy}=\overline{\left|\frac{I_{x,y}}{M_{x,y}}-1\right|}
    \end{equation}
    where $I_{x,y}$ and $M_{x,y}$ are the pixel values at \textit{x,y}
    of the object and model frames, respectively.
    
    In addition to the steps described in detail by vD05 we applied a
    correction for residual noise to the derived tidal parameters.
    In order to do so we created a blank sky frame by aggressively
    masking a fully reduced object image and replacing the masked
    regions with sections from other object frames of the same pixel
    area.
    The resultant blank sky frame conserves the noise characteristics
    of the object images, consisting of both residual large-scale
    flattening variations and pixel scale photon noise.
    Each galaxy model was then added to the blank sky frame
    and a tidal parameter was re-derived for the model galaxy.
    The corrected value is then given by the square-root of the
    difference between the squared tidal-parameter values:
    
    \begin{equation}
      T_c=\sqrt{T^2_{galaxy}-T^2_{model}}
    \end{equation}
    
    By using this correction method we allow the model to grow
    to arbitrarily large radii without erroneously increasing the
    tidal parameter value.
    This is especially useful when a tidal feature exists relatively
    far from the center of its host galaxy but is clearly related to
    it (e.g. NGC 596).
    Figure \ref{fig:tp_clust} demonstrates the effectiveness of this
    method which noticeably improves tidal feature parameterization in
    some galaxies.

    The quantitative tidal parameter is a useful tool for studying the
    evolution of interacting galaxies.
    Not only does it correlate well with visually identified tidal
    disturbances, it also reveals morphological deviations from a
    smooth stellar body in visually less clear cases.
    Such an objective classification is important for comparison with
    similar surveys as well as for studying how stellar disturbances
    correlate with other galactic properties.    
    
    \begin{figure}
      \includegraphics[width=0.47\textwidth]{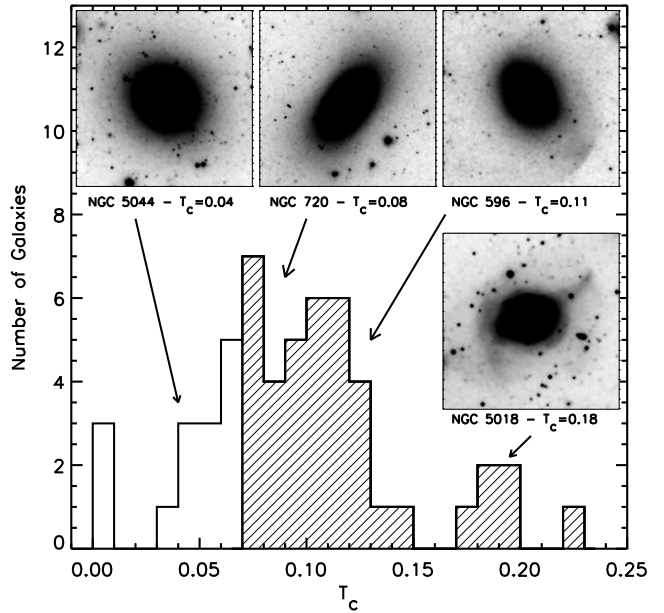}
      \caption{The distribution of derived tidal parameter values. The
      shaded area represents galaxies with a T$_c$ value greater than
      the detection threshold. The four subset images are typical
      examples of various interaction levels.}
      \hfill
      \label{fig:tp_distr}
    \end{figure}
   
  \subsection{Frequency of Tidal Features}
    The sample of 55 galaxies can be divided into three categories
    based on their tidal parameter values.
    In the first group we include galaxies with a corrected tidal
    parameter larger than 0.09.
    These systems all show clear signs of morphological disturbances
    both in the total-counts image and in the model divided 
    frame and they account for roughly 53\% of the sample (29
    galaxies).
    The second category consists of 11 galaxies ($\sim$20\%) with
    $0.07 < T_c < 0.09$ and it includes systems that have
    marginally detected disturbances in their stellar morphologies.
    Galaxies in this group typically show no obvious signs of past
    interactions in the total-counts image but exhibit some
    irregularities in the model divided frame.
    The last sub-group contains 15 galaxies ($\sim$27\%) that lack
    clear interaction signatures in either the total-counts image or
    the model divided frame.
    All galaxies in this group have a tidal parameter value of
    less than 0.07.
    
    Taking $T_c=0.07$ as the threshold, 40 galaxies show morphological
    signs of past interactions in their stellar bodies, accounting for
    $\sim$73\% of the sample (figure \ref{fig:tp_distr}). 
    This result confirms the findings by vD05 and further generalizes
    the phenomenon due to the statistical completeness of this study.
      
    An additional sub-population of the sample consists of galaxies
    that have a projected close companion.
    We identify 7 such pairs, all showing morphological disturbances
    in both galaxies.
    A visual inspection of the companions finds that 5 of them
    have spiral features and are likely disk-systems while
    the other 3  are probably early-type or dwarf galaxies
    with no significant disk component.
    In addition to these interacting systems we identify one pair (NGC
    5638) that shows no interaction signatures in either of the
    galaxies and is appropriately assigned a tidal parameter value of
    0.04.
    The pair fraction is similar to that found by vD05 and although
    the fraction of early-type companions is lower
    \citep[see][]{whitaker_hubble_2008}, the difference is insignificant.
    
    \begin{figure}
      \includegraphics[width=0.47\textwidth]{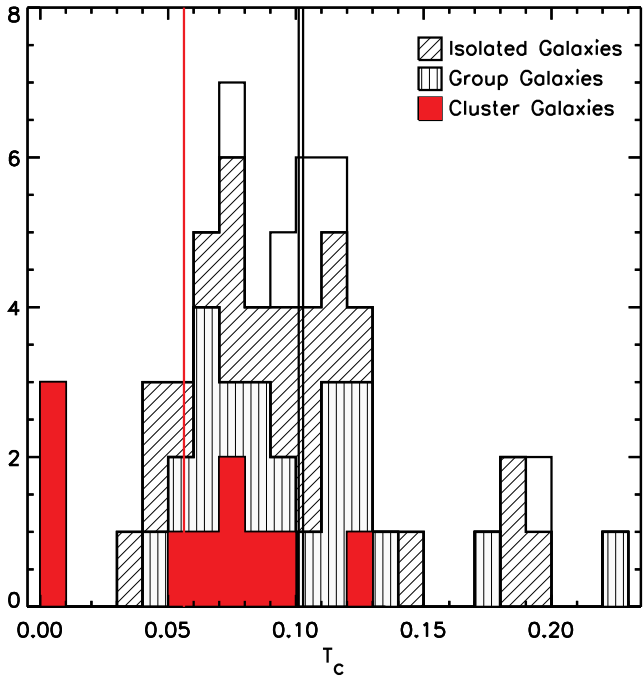}
      \caption{The distribution of tidal parameter values in respect
      to galactic environments. The red histogram represents
      cluster galaxies and the vertical lines show the mean $T_c$
      values of the different populations. Cluster galaxies
      ($\overline{T_{c}}=0.056$) are on average less tidally disturbed than
      group and field galaxies ($\overline{T_{c}}=0.102$).}
      \hfill
      \label{fig:tp_clust2}
    \end{figure}

\section{Properties of Tidally Disturbed Galaxies}
  The derived tidal parameter allows us to quantify the disturbance
  caused to the stellar body of galaxies due to gravitational
  interactions.
  One of the key advantages of such a numerical parameter is that
  it allows us to correlate gravitational disturbance signatures with
  other galactic properties.
  In this section we compare the derived $T_c$ of each galaxy with
  galactic environment, its broadband optical colors and its nuclear
  activity.
  These properties have all been hypothesized to be related to
  gravitational interactions and the subsequent evolution of these
  galaxies.
  
  \subsection{Galaxy Environment\label{galenv}}
    The fraction of galaxies with $T_C \ge 0.07$ in our sample
    increases to $\sim$80\% when cluster members are excluded.
    This is depicted in figure \ref{fig:tp_clust2}, which shows the
    relation between the environment in which the galaxies reside
    and their tidal parameter values.
    A one-sided Mann-Whitney test finds a 1.4\% chance that cluster
    galaxies have a tidal parameter value larger than the sample
    median.

    Interaction signatures seem therefore to be typical of group and
    field galaxies, and relatively rare in clusters.
    The absence of tidal features in clusters might have been
    expected given that mergers are rare in relaxed clusters
    \citep{makino_merger_1997, van_dokkum_high_1999}; on the other
    hand, "galaxy harassment" due to frequent long-range interactions
    has been shown to create long, faint tidal
    tails associated with infalling galaxies
    \citep[e.g.][]{moore_galaxy_1996,moore_morphological_1998}. 
    
  \subsection{Broadband Optical Colors\label{galcol}}
    The relation between the optical colors of elliptical galaxies
    and their absolute magnitude has been known for more than three
    decades \citep[e.g.][]{sandage_color-absolute_1978,
      frogel_photometric_1978}.
    This relation states that giant ellipticals have
    systematically redder colors than their less massive
    counterparts.
    In recent years this observed phenomenon has been mostly
    attributed to the evolution of these galaxies through
    hierarchical growth and mass accretion, thus differentiating it
    from monolithic collapse models.
    This change of paradigm has been supported by both observations
    and numerical simulations of galactic evolution
    \citep[e.g.][]{kauffmann_k-band_1998,
      dokkum_color-magnitude_1998, shioya_tightness_1998}.
    
    In their paper, \cite{schweizer_correlations_1992} show that
    the observed ``fine structure'' value of ellipticals is also
    correlated with the systemic colors of these galaxies.
    The correlation persists even after the observed values are
    normalized by the known color-magnitude relation.
    We perform a similar analysis of our data using normalized
    optical colors in conjunction with the objectively derived tidal
    parameter (see subsection \ref{tpderiv} for details).
    
\begin{figure}
      \includegraphics[width=0.47\textwidth]{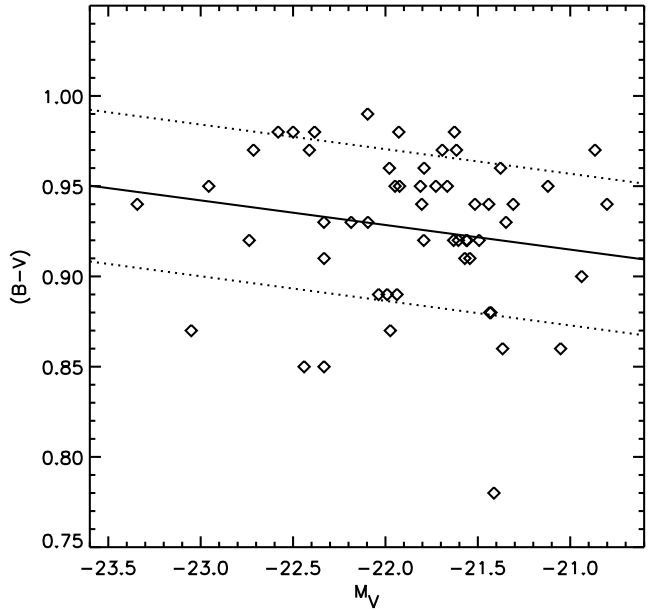}
      \caption{$B-V$ color-magnitude relation of the sample
	galaxies. The solid line is a fit to the data and the dotted
	lines represent 1$\sigma$ deviations from the fit.}
      \hfill
      \label{fig:col_mag}
    \end{figure}

\begin{figure}
      \includegraphics[width=0.47\textwidth]{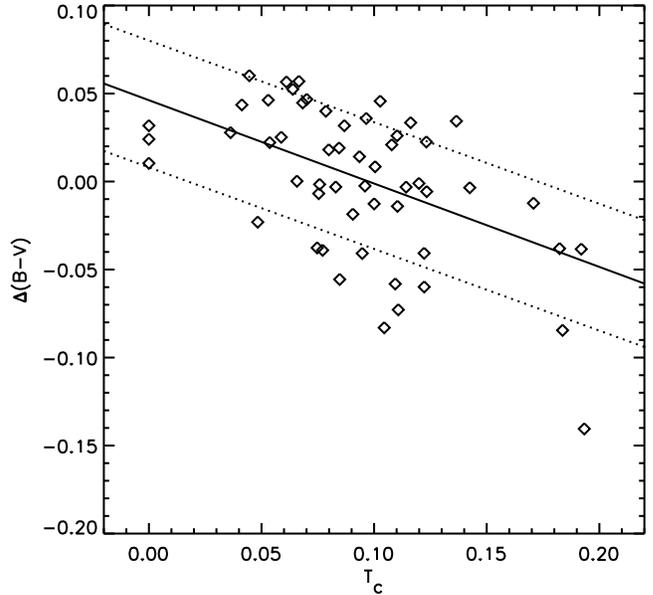}
      \caption{The B-V color is	normalized by the color-magnitude
	relation from figure \ref{fig:col_mag} and plotted against
	the corrected tidal parameter.
	There exists a moderate correlation between tidal disturbances
	and the broadband colors of luminous elliptical galaxies. The
	solid line is a fit to the data and the dotted lines represent
	1$\sigma$ deviations from the fit.}
      \hfill
      \label{fig:tp_colors}
    \end{figure}

    Figure \ref{fig:col_mag} shows the color-magnitude relation of the
    55 sample galaxies for the B-V color, plotted against 
    absolute V magnitudes.
    Color measurements were taken from \cite{michard_near_2005}
    where possible (41 galaxies) and from
    \cite{de_vaucouleurs_third_1991}.
    This relation has an inherit 1$\sigma$ scatter of 0.042 in color
    which cannot be all attributed to errors in measurement and it
    agrees with the slope found by
    \cite{schweizer_correlations_1992}.
    We subtract the color-magnitude relation from the measured colors
    and plot the residuals against the corrected tidal parameter
    values of the galaxies.
    Figure \ref{fig:tp_colors} shows that tidally disturbed galaxies
    are on average bluer than their non-interacting counterparts, in
    agreement with the findings of
    \cite{schweizer_correlations_1992}.
    This correlation is moderate, showing a difference of only
    0.05-0.1 in color between galaxies with tidal parameter values of
    0.0 and 0.2.
    A Spearman rank test finds that the probability that this
    correlation was drawn by chance is less than 0.01\%.

    The correlation between tidal disturbances and broadband colors
    suggests that these merger events probably trigger some star
    formation in the interacting galaxies.
    Alternatively, it is possible that galaxies accreted into
    ellipticals typically contain a significant fraction of young or
    metal-poor stars.
    Such a scenario is not uncommon in groups, where a single
    massive elliptical is surrounded by multiple low mass late-type
    galaxies.
    Nevertheless, it is important to note that all the galaxies in the
    sample are red and that any contribution to their broadband colors
    from star-formation or young stellar populations is small.

  \subsection{Tidal Features and the AGN Duty Cycle\label{galagn}}
    It has been suggested by several authors that gravitational
    interactions play an important role in triggering galactic
    nuclear activity.
    Some of these studies show a correlation between the brightness
    of elliptical galaxies in radio continuum observations and
    morphological disturbances \citep[e.g.][]{heckman_galaxy_1986,
      smith_multicolor_1989}.
    In this scenario minor merger events induce dust and gas
    accretion onto the galactic center and then trigger nuclear
    activity.
    The same mergers should in principal leave their mark on the
    stellar body of the accreting elliptical in the form of
    morphological disturbances such as described in subsection
    \ref{mordist}.
      
\begin{figure}
      \includegraphics[width=0.47\textwidth]{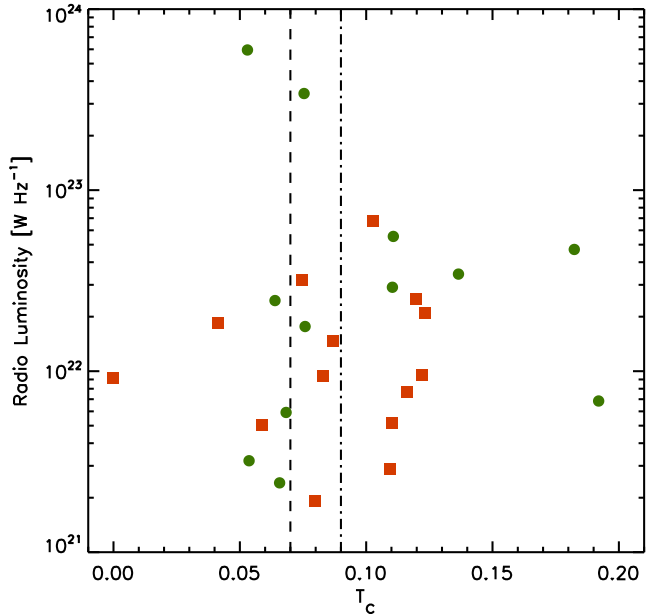}
      \caption{The relation between radio continuum flux and tidal
	parameter for 26 of the sample galaxies.
	The existence of radio-load, non-interacting galaxies suggests
	that gravitational interactions are not the only, and perhaps
        not the most important, AGN triggering mechanism.}
      \hfill
      \label{fig:radiotp}
    \end{figure}

    In order to test this hypothesis we compiled a list of 6cm and
    20cm flux measurements for 26 galaxies with publicly available
    data.
    The data were collected form a number of catalogs, including the
    NRAO VLA Sky Survey \citep{condon_nrao_1998}, the FIRST survey
    \citep{white_catalog_1997}, the NORTH6CM and the NORTH20CM
    databases \citep{becker_new_1991,white_new_1992}, the Green-Bank
    6cm survey \citep{gregory_gb6_1996} and the Dixon Master List of
    Radio Sources \citep{dixon_master_1970}.
    As not all galaxies have published measurements for
    both wavelengths we derive a correction factor from the 20cm to
    6cm relation using 8 galaxies for which both data are
    available.
    This relation is expressed by the following equation:
    \begin{equation}
      \log{F_{6{\rm cm}}} = 1.20+0.53\log{F_{20{\rm cm}}}
      \label{eq:radio20to6}
    \end{equation}
    where $F_{6m}$ and $F_{20m}$ are the measured fluxes in the 6cm
    and 20cm bands, respectively.
    The 1$\sigma$ scatter in this relation is 0.29 dex.
    Figure \ref{fig:radiotp} shows the correlation between radio
    continuum flux and tidal parameter for the sample galaxies.
    Plotted values are either 6cm observations or 20cm observations
    corrected to 6cm using the linear relation from equation
    \ref{eq:radio20to6}.

    We see no relation between the two parameters.
    This lack of a tight correlation is itself not surprising given
    the very different expected lifetimes of tidal features and the
    radio duty cycle.
    Tidal interactions, as portrayed by their assigned tidal parameter
    values, are expected to survive for a long time in the outskirts
    of elliptical galaxies.
    This means that a disturbed galaxy can retain its large value of
    tidal parameter even after the nuclear activity has quieted.

    This can be a viable explanation for the scatter at high values of
    $T_c$ but not for that at the low end of the $T_c$ distribution.
    Since the lifetime of the radio mode is only roughly 10$^8$
    years \citep[e.g.,][]{croton_many_2006,shabala_duty_2008},
    galaxies with a low T$_C$ value are all expected to be 
    quiet at radio wavelength.
    The existence of radio-loud AGN in undisturbed ellipticals is very
    interesting and may imply that gravitational interactions are not
    the only and possibly not the most important AGN triggering
    mechanism.
    
\section{Accretion Rate and Inferred Mass Growth\label{accrate}}
  In the following section we will attempt to estimate the
  rate at which ellipticals grow through mergers using two techniques.
  First we estimate the survival time of the observed tidal
  features using the dynamical time of the system at the innermost
  radius of feature detection.
  This estimate relies on the assumption that tidal features
  dissipate due to dynamical mixing in their environment.
  As an alternative approach we analyze the interacting pairs in
  the sample using friction timescale approximations as suggested
  by, e.g., \cite{patton_new_2000}.
  
  \subsection{Typical Tidal Feature Survival Time\label{orbtime}}
    Most of the merger rate estimates in the literature measure the
    frequency and properties of close interacting pairs.
    In this work, however, we are able to use an alternative approach.
    For each galaxy we measure the smallest radius where tidal
    features are seen and estimate the age of the merger event from
    the dynamical time at that radius.
      
\begin{figure}
      \includegraphics[width=0.47\textwidth]{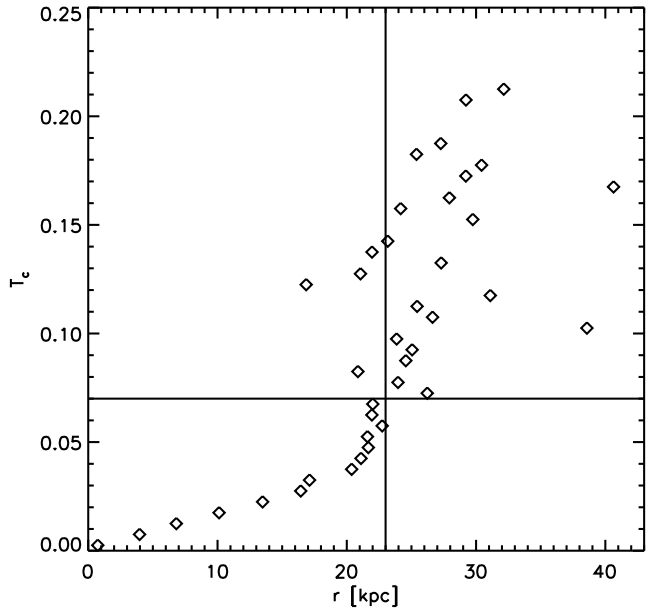}
      \caption{The average distance at which tidal features are
      detected. The derived fractional tidal parameter values are
      binned with bin size of 0.005. The horizontal line marks our
      detection threshold of T$_c$=0.07. Most detections are to the
      right of the 23 kpc line which is the average at the 0.07
      threshold.}
      \hfill
      \label{fig:tprad3}
    \end{figure}

    The dynamical time at a distance $r$ from the center of a galaxy
    with circular velocity $v_c(r)$ is defined as the time it would 
    take a test particle to complete one-quarter of an orbit under the
    gravitational potential of a homogeneous sphere.
    We use this to assess the survival time of a tidal feature,
    assuming that after completing $N_o$ orbits the feature will
    dissolve into the smooth stellar distribution due to the inherent
    velocity dispersion and differential rotation of the galactic
    potential.
    The survival time is then given by:
    \begin{equation}
      T_{s}=N_o T_{dyn}=N_o\frac{\pi}{2}\frac{r}{v_c(r_{min})}
    \end{equation}
    where $N_o$ is the number of dynamical times it takes tidal
    features to dissolve into the stellar body of the galaxy and
    $r_{min}$ is the smallest radius at which we detect tidal
    features.
    In order to find this radius we derived fractional tidal
    parameter values as a function of radius for the sample galaxies,
    thus measuring tidal distortions within limiting radii.
    Figure \ref{fig:tprad3} shows the average radius of each
    fractional $T_c$ bin of size $\Delta T_c=0.005$.
    The majority of the significant tidal features lie at radii
    larger than 23 kpc which is also the average distance of features
    with $T_c\sim 0.07$.
    We estimate $v_c$ by measuring the circular velocity of a
    model galaxy with an NFW dark matter profile
    \citep{navarro_universal_1997} and a Hernquist stellar profile
    \citep{hernquist_analytical_1990} at the same radius.    
    The typical time since the last accretion event for
    these galaxies is then:
    \begin{equation}
      T_{s}=0.35 N_o \left(\frac{r}{23\ {\rm kpc}}\right)
      \left(\frac{395\ {\rm km\ s^{-1}}}{v}\right){\rm Gyr},
      \label{eq:torb}
    \end{equation}

    The actual survival time, however, can vary greatly based on the
    tidal feature morphology and kinematics that result from different
    interactions and collisions.
    Since typical stellar orbits in elliptical galaxies are highly
    eccentric and have significant precession, radial features will
    dissolve much quicker than tangential ones.
    For example, a radial linear feature is only expected to be
    detectable for about one dynamical time ($N_o \sim 1$), whereas
    shell-like features are expected to survive for many ($N_o>10$).
    
    This technique of estimating the interaction time-scale of nearby
    ellipticals is important as it does not rely on close pairs, which
    are much less common.
    Nevertheless, when deriving the survival time we did not treat any
    biases that might be caused by the projection of tidal features
    onto the plane of the sky.
    For example, a stellar ring around a galaxy at some distance $r$
    may appear as a radial feature extending all the way through the
    center of the galaxy if aligned properly.
    Rather than individually simulating all sample galaxies we assume
    that feature orientations are random and that our sample is
    sufficiently large to minimize their effect on this result.
    
  \subsection{Implied Merger and Mass Growth Rate}
    Assuming that in a given merger event all the mass from both
    galaxies is retained in the system we can write the mass growth
    rate of an accreting elliptical:
    \begin{equation}
      \frac{dM}{M} = \frac{m_2}{m_1}\frac{f_T}{T_s} dt
      \label{eq:growth}
    \end{equation}
    where $m_2/m_1$ is the progenitor galaxy mass ratio,
    $T_S$ is the tidal feature survival time (see equation
    \ref{eq:torb}) and $f_T$ is the fraction of observed galaxies with
    tidal distortions.    
    In this paper we confirm the value of $f_T\sim0.7$ and measure
    $T_{s}(r_{min})\sim 0.35N_o$ Gyrs.
    
    Two significant uncertainties remain in equation
    \ref{eq:growth}.
    Both the progenitor mass ratio $m_2/m_1$ and the number of
    survival orbits $N_o$ cannot be observed directly from our data
    and require simulations of multiple systems in order to constrain
    their values.
    In addition, both parameters can likely have a wide range of
    values, varying between systems and tidal feature morphologies.
    We therefore adopt conservative values for these parameters to get
    an estimate for the mass growth rate due to gravitational
    interactions:
    \begin{equation}
      \frac{dM}{M} = 0.2 \left(\frac{m_2/m_1}{0.1}\right)
      \left(\frac{f_t}{0.7}\right) \left(\frac{1}{N_o}\right)
      \frac{dT}{{\rm Gyr}}
    \end{equation}

    Assuming a constant rate since $z\sim 2$, this implies that an
    isolated galaxy can grow up to 3 times its initial collapsed mass
    by minor mergers only.
    This result is in good agreement with recent studies that show
    that galaxies can grow significantly in mass through minor mergers
    \citep{naab_formation_2007, bournaud_multiple_2007,
      bezanson_relation_2009}.
    
  \subsection{Pair Friction Time\label{pftime}}
    For purposes of comparison with previous studies we perform a
    second analysis of the interaction timescale using a subsample
    of close pairs of interacting galaxies.
    In these systems the time it would take the two galaxies to
    merge can by approximated by the dynamical friction timescale
    \citep{patton_new_2000}:
    
    \begin{equation}
      T_{fric} = \frac{2.64\pow10{5}r^2 v_c}{M \ln{\Lambda}}\ {\rm
        Gyr},
      \label{tfric}
    \end{equation}
    where $r$ is the physical separation of the pairs, $v_c$ is the
    circular velocity, $M$ is the mass of the secondary galaxy and
    $\Lambda$ is the usual Coulomb logarithm
    \citep{binney_galactic_1987}.
    In our galaxy sample we identify 7 merging pairs with a mean
    projected distance of 21.4$\pm$6.7 kpc, resulting in a
    circular velocity of $\sim$400\kms.
    Using typical values of M$_{c}=7\pow10{10}$M$_{\odot}$ and
    $\ln\Lambda=2$ we get a second interaction time estimate of
    $\sim$0.53 Gyr.    
    This result is in good agreement with vD05 and it is similar to
    $T_s$ for $N_o\sim2$

\section{Conclusions}
  The evolution and growth of elliptical galaxies in the nearby
  Universe is driven by gravitational interactions and merger
  events.
  The rate of these interactions, which are not accompanied by
  significant star formation, can be estimated from the morphological
  distortions that they cause to the stellar bodies of 
  colliding galaxies.
  vD05 showed that at z$\sim$0.1
  signatures of past interactions are observed in most ellipticals and
  can be accounted for in more than 73\% of the galaxies.
  In this work we confirmed this result upon analyzing 55
  nearby ellipticals in a variety of galactic and environmental
  properties.
  This is the first time that such a study is performed on a
  statistically complete sample and it shows, above all, that
  elliptical galaxies are still assembling via mergers and accretion
  events in the nearby Universe.
  Although this high rate of gravitational interaction signatures is
  not a new finding, the completeness of our sample implies that this
  is an important phenomenon for all ellipticals and that it is not
  restricted to certain populations.
  Moreover, the high frequency of tidal features strongly suggests
  that these interactions play a critical role in the evolution and
  growth of elliptical galaxies.
  
  In this paper we explored the link between tidal disturbances and
  other evolution driven properties of the sample galaxies.
  We found that interacting systems have slightly bluer
  broadband colors than non interacting ones, implying that these
  mergers are accompanied by little or no star formation.
  One explanation for this is that the progenitor galaxies are
  mainly composed of old stellar populations, thus producing a red
  galaxy.
  Alternatively, the total flux from young stars in the accreted
  galaxies is small compared to the accumulated light from old
  stars in the ellipticals, suggesting that these are mainly minor
  merger events.
    
  We also examined the proposed link between gravitational
  interactions and nuclear activity in ellipticals (subsection
  \ref{galagn}).
  We accomplished this by relating the radio continuum flux of the
  sample galaxies to their respective tidal parameter values.
  Given the long dynamical time in the outskirts of ellipticals, tidal
  features, unlike radio activity, are expected to survive there for a
  long time.
  Using equation \ref{eq:torb} it is easy to show that at a radius of
  50kpc a tidal feature will remain for $\sim10^9$ years.
  On the other hand, the standard assumptions predict a lifetime of
  only $10^8$ years for the AGN radio mode.
  This suggests that all radio loud galaxies are expected to have a
  high value of tidal parameter, in contradiction to our findings
  (figure \ref{fig:radiotp}).
  We conclude that gravitational interactions are therefore an
  unlikely candidate for provoking nuclear activity in elliptical
  galaxies.

  Lastly, we estimated the lifetime of observed tidal features by
  measuring the dynamical time of the galaxies at the innermost radius
  of feature detection.
  We found that the sample galaxies are generally relaxed at radii
  smaller than $\sim$23 kpc, implying that nearby ellipticals typically
  experience gravitational interactions every $\sim$0.35 Gyrs.
  This value yields a mass accretion rate of dM/M$\sim$0.2 with large
  uncertainty.
  The derived rate of mass growth therefore shows that elliptical
  galaxies can grow significantly through (presumably mostly minor)
  mergers and low-mass accretion events.

\appendix
  In this section we present the full data atlas. All the images were
  smoothed using a box-car kernel of dimensions 8''$\times$8'' and
  were further bin by a factor 2$\times$2 in order to bring out some
  of the fainter tidal features. Similar to figure
  \ref{fig:atlas_sub}, every line shows from left to right the full
  data set, object masked frame, the fitted model and the
  model-divided residual image.
  The corrected tidal parameter and a scale bar are also shown.
  \\
  \\
  The full paper and appendix are available online at:
  http://www.astro.yale.edu/obey

\bibliography{ms}			

\begin{thebibliography}{55}
\expandafter\ifx\csname natexlab\endcsname\relax\def\natexlab#1{#1}\fi

\bibitem[{Becker {et~al.}(1991)Becker, White, \& Edwards}]{becker_new_1991}
Becker, R.~H., White, R.~L., \& Edwards, A.~L. 1991, Astrophysical Journal
  Supplement Series, 75, 1

\bibitem[{Bertin \& Arnouts(1996)}]{bertin_sextractor:_1996}
Bertin, E., \& Arnouts, S. 1996, Astronomy and Astrophysics Supplement Series,
  117, 393

\bibitem[{Bezanson {et~al.}(2009)Bezanson, van Dokkum, Tal, Marchesini, Kriek,
  Franx, \& Coppi}]{bezanson_relation_2009}
Bezanson, R., van Dokkum, P.~G., Tal, T., Marchesini, D., Kriek, M., Franx, M.,
  \& Coppi, P. 2009, Astrophysical Journal, 697, 1290

\bibitem[{Binney \& Tremaine(1987)}]{binney_galactic_1987}
Binney, J., \& Tremaine, S. 1987, Galactic dynamics (Princeton)

\bibitem[{Blanton {et~al.}(2003)Blanton, Hogg, Bahcall, Brinkmann, Britton,
  Connolly, Csabai, Fukugita, Loveday, Meiksin, Munn, Nichol, Okamura, Quinn,
  Schneider, Shimasaku, Strauss, Tegmark, Vogeley, \&
  Weinberg}]{blanton_galaxy_2003}
Blanton, M.~R., Hogg, D.~W., Bahcall, N.~A., Brinkmann, J., Britton, M.,
  Connolly, A.~J., Csabai, I., Fukugita, M., Loveday, J., Meiksin, A., Munn,
  J.~A., Nichol, R.~C., Okamura, S., Quinn, T., Schneider, D.~P., Shimasaku,
  K., Strauss, M.~A., Tegmark, M., Vogeley, M.~S., \& Weinberg, D.~H. 2003,
  Astrophysical Journal, 592, 819

\bibitem[{Bournaud {et~al.}(2007)Bournaud, Jog, \&
  Combes}]{bournaud_multiple_2007}
Bournaud, F., Jog, C.~J., \& Combes, F. 2007, Astronomy and Astrophysics, 476,
  1179

\bibitem[{Bower {et~al.}(1992)Bower, Lucey, \& Ellis}]{bower_precision_1992}
Bower, R.~G., Lucey, J.~R., \& Ellis, R.~S. 1992, Monthly Notices of the Royal
  Astronomical Society, 254, 589

\bibitem[{{Boylan-Kolchin} {et~al.}(2006){Boylan-Kolchin}, Ma, \&
  Quataert}]{boylan-kolchin_red_2006}
{Boylan-Kolchin}, M., Ma, C., \& Quataert, E. 2006, Monthly Notices of the
  Royal Astronomical Society, 369, 1081

\bibitem[{Chang {et~al.}(2006)Chang, Gallazzi, Kauffmann, Charlot, Ivezic,
  Brinchmann, \& Heckman}]{chang_colours_2006}
Chang, R., Gallazzi, A., Kauffmann, G., Charlot, S., Ivezic, Z., Brinchmann,
  J., \& Heckman, T.~M. 2006, Monthly Notices of the Royal Astronomical
  Society, 366, 717

\bibitem[{Colbert {et~al.}(2001)Colbert, Mulchaey, \&
  Zabludoff}]{colbert_optical_2001}
Colbert, J.~W., Mulchaey, J.~S., \& Zabludoff, A.~I. 2001, Astronomical
  Journal, 121, 808

\bibitem[{Combes {et~al.}(1995)Combes, Rampazzo, Bonfanti, Pringniel, \&
  Sulentic}]{combes_stellar_1995}
Combes, F., Rampazzo, R., Bonfanti, P.~P., Pringniel, P., \& Sulentic, J.~W.
  1995, Astronomy and Astrophysics, 297, 37

\bibitem[{Condon {et~al.}(1998)Condon, Cotton, Greisen, Yin, Perley, Taylor, \&
  Broderick}]{condon_nrao_1998}
Condon, J.~J., Cotton, W.~D., Greisen, E.~W., Yin, Q.~F., Perley, R.~A.,
  Taylor, G.~B., \& Broderick, J.~J. 1998, Astronomical Journal, 115, 1693

\bibitem[{Croton {et~al.}(2006)Croton, Springel, White, Lucia, Frenk, Gao,
  Jenkins, Kauffmann, Navarro, \& Yoshida}]{croton_many_2006}
Croton, D.~J., Springel, V., White, S. D.~M., Lucia, G.~D., Frenk, C.~S., Gao,
  L., Jenkins, A., Kauffmann, G., Navarro, J.~F., \& Yoshida, N. 2006, Monthly
  Notices of the Royal Astronomical Society, 365, 11

\bibitem[{de~Vaucouleurs {et~al.}(1991)de~Vaucouleurs, de~Vaucouleurs, Corwin,
  Buta, Paturel, \& Fouque}]{de_vaucouleurs_third_1991}
de~Vaucouleurs, G., de~Vaucouleurs, A., Corwin, H.~G., Buta, R.~J., Paturel,
  G., \& Fouque, P. 1991, Third Reference Catalogue of Bright Galaxies
  ({Springer-Verlag} Berlin Heidelberg New York)

\bibitem[{Dekel \& Birnboim(2008)}]{dekel_gravitational_2008}
Dekel, A., \& Birnboim, Y. 2008, Monthly Notices of the Royal Astronomical
  Society, 383, 119

\bibitem[{Dixon(1970)}]{dixon_master_1970}
Dixon, R.~S. 1970, Astrophysical Journal Supplement Series, 20, 1

\bibitem[{Dupraz \& Combes(1986)}]{dupraz_shells_1986}
Dupraz, C., \& Combes, F. 1986, Astronomy and Astrophysics, 166, 53

\bibitem[{Feldmann {et~al.}(2008)Feldmann, Mayer, \&
  Carollo}]{feldmann_tidal_2008}
Feldmann, R., Mayer, L., \& Carollo, C.~M. 2008, Astrophysical Journal, 684,
  1062

\bibitem[{Frogel {et~al.}(1978)Frogel, Persson, Matthews, \&
  Aaronson}]{frogel_photometric_1978}
Frogel, J.~A., Persson, S.~E., Matthews, K., \& Aaronson, M. 1978,
  Astrophysical Journal, 220, 75

\bibitem[{Gregory {et~al.}(1996)Gregory, Scott, Douglas, \&
  Condon}]{gregory_gb6_1996}
Gregory, P.~C., Scott, W.~K., Douglas, K., \& Condon, J.~J. 1996, Astrophysical
  Journal Supplement Series, 103, 427

\bibitem[{Heckman {et~al.}(1986)Heckman, Smith, Baum, van Breugel, Miley,
  Illingworth, Bothun, \& Balick}]{heckman_galaxy_1986}
Heckman, T.~M., Smith, E.~P., Baum, S.~A., van Breugel, W. J.~M., Miley, G.~K.,
  Illingworth, G.~D., Bothun, G.~D., \& Balick, B. 1986, Astrophysical Journal,
  311, 526

\bibitem[{Hernquist(1990)}]{hernquist_analytical_1990}
Hernquist, L. 1990, Astrophysical Journal, 356, 359

\bibitem[{Kauffmann(1996)}]{kauffmann_age_1996}
Kauffmann, G. 1996, Monthly Notices of the Royal Astronomical Society, 281, 487

\bibitem[{Kauffmann \& Charlot(1998)}]{kauffmann_k-band_1998}
Kauffmann, G., \& Charlot, S. 1998, Monthly Notices of the Royal Astronomical
  Society, 297, L23

\bibitem[{Kauffmann {et~al.}(1993)Kauffmann, White, \&
  Guiderdoni}]{kauffmann_formation_1993}
Kauffmann, G., White, S. D.~M., \& Guiderdoni, B. 1993, Monthly Notices of the
  Royal Astronomical Society, 264, 201

\bibitem[{Kenney {et~al.}(2008)Kenney, Tal, Crowl, Feldmeier, \&
  Jacoby}]{kenney_spectacular_2008}
Kenney, J. D.~P., Tal, T., Crowl, H.~H., Feldmeier, J., \& Jacoby, G.~H. 2008,
  Astrophysical Journal, 687, L69

\bibitem[{Knapp {et~al.}(1985)Knapp, Turner, \&
  Cunniffe}]{knapp_statistical_1985}
Knapp, G.~R., Turner, E.~L., \& Cunniffe, P.~E. 1985, Astronomical Journal, 90,
  454

\bibitem[{Kormendy(1984)}]{kormendy_recognizing_1984}
Kormendy, J. 1984, Astrophysical Journal, 287, 577

\bibitem[{Makino \& Hut(1997)}]{makino_merger_1997}
Makino, J., \& Hut, P. 1997, Astrophysical Journal, 481, 83

\bibitem[{{McIntosh} {et~al.}(2008){McIntosh}, Guo, Hertzberg, Katz, Mo,
  van~den Bosch, \& Yang}]{mcintosh_ongoing_2008}
{McIntosh}, D.~H., Guo, Y., Hertzberg, J., Katz, N., Mo, H.~J., van~den Bosch,
  F.~C., \& Yang, X. 2008, Monthly Notices of the Royal Astronomical Society,
  388, 1537

\bibitem[{Michard(2005)}]{michard_near_2005}
Michard, R. 2005, Astronomy and Astrophysics, 441, 451

\bibitem[{Moore {et~al.}(1996)Moore, Katz, Lake, Dressler, \&
  Oemler}]{moore_galaxy_1996}
Moore, B., Katz, N., Lake, G., Dressler, A., \& Oemler, A. 1996, Nature, 379,
  613

\bibitem[{Moore {et~al.}(1998)Moore, Lake, \& Katz}]{moore_morphological_1998}
Moore, B., Lake, G., \& Katz, N. 1998, Astrophysical Journal, 495, 139

\bibitem[{Naab {et~al.}(2007)Naab, Johansson, Ostriker, \&
  Efstathiou}]{naab_formation_2007}
Naab, T., Johansson, P.~H., Ostriker, J.~P., \& Efstathiou, G. 2007,
  Astrophysical Journal, 658, 710

\bibitem[{Naab {et~al.}(2006)Naab, Khochfar, \& Burkert}]{naab_properties_2006}
Naab, T., Khochfar, S., \& Burkert, A. 2006, Astrophysical Journal, 636, L81

\bibitem[{Navarro {et~al.}(1997)Navarro, Frenk, \&
  White}]{navarro_universal_1997}
Navarro, J.~F., Frenk, C.~S., \& White, S. D.~M. 1997, Astrophysical Journal,
  490, 493

\bibitem[{Patton {et~al.}(2000)Patton, Carlberg, Marzke, Pritchet, da~Costa, \&
  Pellegrini}]{patton_new_2000}
Patton, D.~R., Carlberg, R.~G., Marzke, R.~O., Pritchet, C.~J., da~Costa,
  L.~N., \& Pellegrini, P.~S. 2000, Astrophysical Journal, 536, 153

\bibitem[{Prugniel \& Heraudeau(1998)}]{prugniel_general_1998}
Prugniel, P., \& Heraudeau, P. 1998, {VizieR} Online Data Catalog, 7206, 0

\bibitem[{Sadler(2001)}]{sadler_hi_2001}
Sadler, E.~M. 2001, in Gas and Galaxy Evolution, Vol. 240, 445

\bibitem[{Sandage \& Visvanathan(1978)}]{sandage_color-absolute_1978}
Sandage, A., \& Visvanathan, N. 1978, Astrophysical Journal, 223, 707

\bibitem[{Sanders(1980)}]{sanders_neutral_1980}
Sanders, R.~H. 1980, Astrophysical Journal, 242, 931

\bibitem[{Schlegel {et~al.}(1998)Schlegel, Finkbeiner, \&
  Davis}]{schlegel_maps_1998}
Schlegel, D.~J., Finkbeiner, D.~P., \& Davis, M. 1998, Astrophysical Journal,
  500, 525

\bibitem[{Schweizer \& Seitzer(1992)}]{schweizer_correlations_1992}
Schweizer, F., \& Seitzer, P. 1992, Astronomical Journal, 104, 1039

\bibitem[{Shabala {et~al.}(2008)Shabala, Ash, Alexander, \&
  Riley}]{shabala_duty_2008}
Shabala, S.~S., Ash, S., Alexander, P., \& Riley, J.~M. 2008, Monthly Notices
  of the Royal Astronomical Society, 388, 625

\bibitem[{Shioya \& Bekki(1998)}]{shioya_tightness_1998}
Shioya, Y., \& Bekki, K. 1998, Astrophysical Journal, 504, 42

\bibitem[{Smith \& Heckman(1989)}]{smith_multicolor_1989}
Smith, E.~P., \& Heckman, T.~M. 1989, Astrophysical Journal, 341, 658

\bibitem[{Tully(1988)}]{tully_nearby_1988}
Tully, R.~B. 1988, Nearby galaxies catalog (Cambridge and New York, Cambridge
  University Press), 221

\bibitem[{van Dokkum(2005)}]{dokkum_recent_2005}
van Dokkum, P.~G. 2005, Astronomical Journal, 130, 2647

\bibitem[{van Dokkum {et~al.}(1999)van Dokkum, Franx, Fabricant, Kelson, \&
  Illingworth}]{van_dokkum_high_1999}
van Dokkum, P.~G., Franx, M., Fabricant, D., Kelson, D.~D., \& Illingworth,
  G.~D. 1999, Astrophysical Journal, 520, L95

\bibitem[{van Dokkum {et~al.}(1998)van Dokkum, Franx, Kelson, Illingworth,
  Fisher, \& Fabricant}]{dokkum_color-magnitude_1998}
van Dokkum, P.~G., Franx, M., Kelson, D.~D., Illingworth, G.~D., Fisher, D., \&
  Fabricant, D. 1998, Astrophysical Journal, 500, 714

\bibitem[{Visvanathan \& Sandage(1977)}]{visvanathan_color-absolute_1977}
Visvanathan, N., \& Sandage, A. 1977, Astrophysical Journal, 216, 214

\bibitem[{Whitaker \& van Dokkum(2008)}]{whitaker_hubble_2008}
Whitaker, K.~E., \& van Dokkum, P.~G. 2008, Astrophysical Journal, 676, L105

\bibitem[{White \& Becker(1992)}]{white_new_1992}
White, R.~L., \& Becker, R.~H. 1992, Astrophysical Journal Supplement Series,
  79, 331

\bibitem[{White {et~al.}(1997)White, Becker, Helfand, \&
  Gregg}]{white_catalog_1997}
White, R.~L., Becker, R.~H., Helfand, D.~J., \& Gregg, M.~D. 1997,
  Astrophysical Journal, 475, 479

\bibitem[{White \& Frenk(1991)}]{white_galaxy_1991}
White, S. D.~M., \& Frenk, C.~S. 1991, Astrophysical Journal, 379, 52

\end{thebibliography}
\bibliographystyle{apj}

\end{document}